\documentclass[letterpaper, 10 pt, conference]{ieeeconf}  

\IEEEoverridecommandlockouts                              
\title{\LARGE \bf
 Event-triggered observer design for linear systems
}

\author{E. Petri,	
	\thanks{} 
	R. Postoyan,
	\thanks{
		 E. Petri and R. Postoyan are with the Universit\'e de Lorraine, CNRS, CRAN, F-54000 Nancy, France.   
		(\texttt{elena.petri@univ-lorraine.fr}, \texttt{romain.postoyan@univ-lorraine.fr}).}
	D. Astolfi,
	\thanks{ D. Astolfi is with  
		Universit\'e Claude Bernard Lyon 1, CNRS, LAGEPP UMR 5007, 43 Boulevard du 11 Novembre 1918, F-69100,
		Villeurbanne, France. (\texttt{daniele.astolfi@univ-lyon1.fr}).}
	D. Ne\v{s}i\'c
	\thanks{ D. Ne\v{s}i\'c is with the Department of Electrical and Electronic Engineering, The University of Melbourne, Parkville, 3010 Victoria Australia.
		(\texttt{dnesic@unimelb.edu.au}). }
	and W.P.M.H. Heemels
	\thanks{ W.P.M.H. Heemels is with the Department of Mechanical Engineering,
		Eindhoven University of Technology, The Netherlands.
		(\texttt{m.heemels@tue.nl}).}
	\thanks{This work was funded by Lorraine Universit\'e d'Excellence LUE, HANDY project ANR-18-CE40-0010-02, the France Australian collaboration project IRP-ARS CNRS and the Australian Research Council under the Discovery Project DP200101303.  }
}

\usepackage{graphicx,fleqn,amstext,amsmath,amssymb,color,psfrag,mathabx,booktabs}
\usepackage[latin1]{inputenc}
\usepackage[colorlinks=true, letterpaper,    
urlcolor=black,     
citecolor=blue,
menucolor=black,    
linkcolor=black,    
pagecolor=black,    
breaklinks=true,
]{hyperref}

\usepackage{amsthm}

\usepackage[noadjust]{cite}

\newcommand{\R}{\ensuremath{\mathbb{R}}}

\newcommand{\Rlo}{\ensuremath{\mathbb{R}_{\geq 0}}}

\newcommand{\Zo}{\ensuremath{\mathbb{Z}_{\geq 0}}}
\newcommand{\Zp}{\ensuremath{\mathbb{Z}_{> 0}}}
\newcommand{\Z}{\ensuremath{\mathbb{Z}}}

\definecolor{bleucit}{rgb}{0.2,0.4,0.6} 

\definecolor{blue_cv}{rgb}{0.09,0.35,0.78}

\newcommand{\dom}{\ensuremath{\text{dom}\,}}

\newcommand{\norm}[1]{\ensuremath{\left\|{#1}\right\|}}

\theoremstyle{theorem}

\newtheorem{ass}{\textnormal{\textbf{Assumption}}}

\newtheorem{lem}{Lemma}

\newtheorem{thm}{\textnormal{\textbf{Theorem}}}

\newtheorem{rem}{\textnormal{\textbf{Remark}}}

\usepackage[noadjust]{cite}

\let\labelindent\relax
\usepackage{enumitem}
\usepackage{tikz}
\usetikzlibrary{shapes,arrows,calc}
\definecolor{MyGreen}{RGB}{50,140,80}

\usepackage[export]{adjustbox}
\usepackage{ circuitikz }
 \usetikzlibrary{arrows}

\begin{document}

\maketitle
\thispagestyle{empty}
\pagestyle{empty}

\begin{abstract}
We present an event-triggered observer design for linear time-invariant systems, where the measured output is sent to the observer only when a triggering condition is satisfied. We proceed by emulation and we first construct a continuous-time Luenberger observer. We then propose a dynamic rule to trigger transmissions, which only depends on the plant output and an auxiliary scalar state variable. The overall system is modeled as a hybrid system, for which a jump corresponds to an output transmission. We show that the proposed event-triggered observer guarantees global practical asymptotic stability for the estimation error dynamics. Moreover, under mild boundedness conditions on the plant state and its input, we prove that there exists a uniform strictly positive minimum inter-event time between any two consecutive transmissions, guaranteeing that the system does not exhibit Zeno solutions. Finally, the proposed approach is applied to a numerical case study of a lithium-ion battery.
\end{abstract}

\section{Introduction}
In many applications, the system state is not directly accessible and needs to be estimated based on the plant input, the measured output and a model of the dynamics using an observer. 
When the sensors and the observer are not co-located, output measurements may need to be transmitted to the observer via a digital network. The transmission policy then has an impact on the convergence speed, robustness of the estimator, as well as on the amount of communication resources required.  
An option is to generate transmissions based on time, in which case we talk of time-triggered strategies for which various results are available in the literature, see, e.g., \cite{postoyan2011framework, li2017robust, ferrante2016state, mazenc2015design}. A possible drawback of this paradigm is that the output measurements are sent over the network even when these are not needed, which can lead to unnecessary resources usage. To overcome this drawback, an alternative is to use event-triggered transmissions.
In this case, an event-based triggering rule monitors the plant measurement and/or the observer state and decides when an output transmission is needed. In this way, it is possible to reduce the number of transmissions over the network, while still ensuring good estimation performance. 

Various works in the literature provide event-triggered estimation schemes. Many papers propose triggering rules to generate the transmission instants, which require a copy of the observer to be implemented with the sensors, see e.g., \cite{li2010event,shi2014event2,li2011performance,trimpe2014stability, scheres2021Event}. This may not be always feasible in applications for which the sensors have limited computation capabilities. 
 An alternative is offered by self-triggering policies where the observer decides when it needs to receive a new output measurement,
 see e.g., \cite{andrieu2015self, rabehi2020finite}, and sends a request to receive new output data. In this case, the plant output is not continuously monitored. 
Another possible solution is to follow the event-triggered approach, without using a local observer and to implement a triggering rule where the sensor decides when to transmit only based on the measured output and its past transmitted value(s), see, e.g., \cite{han2015stochastic, shi2016event, huang2019robust, etienne2017periodic,etienne2017asynchronous,etienne2016event, sijs2012event, sijs2013event}. 
 
In this paper, we adopt this last approach because it keeps monitoring the plant output, which may lead to less transmissions compared to a self-triggered approach, and it does not require a copy of the observer, which simplifies its implementation. The main novelty is a new triggering rule, which involves an auxiliary scalar variable, that has several benefits as explained in the sequel. 
In particular, we present an event-triggered observer for deterministic linear time-invariant continuous-time systems. We follow an emulation-based design in the sense that we first design a Luenberger observer for the continuous-time plant ignoring the packet based nature of communication network. Secondly, we take into account the latter and develop a triggering rule to approximately preserve the original properties of the observer. As already stated, we desire the triggering rule not to rely on a copy of the observer, which might be computational prohibitive. Instead, we only require the sensors to have enough computation resources to run a simple scalar linear filter.
 To be precise, the proposed policy is inspired by dynamic triggering rules used in the event-triggered control literature \cite{girard2014dynamic, tanwani2015using, tabuada2007event} and in \cite{rabehi2020finite}, where self-triggered interval observers are designed. In particular, our strategy consists in filtering an absolute threshold strategy, as opposed to the relative threshold technique as done in the context of control in \cite{girard2014dynamic, tanwani2015using, tabuada2007event}. Indeed, the latter cannot be implemented for estimation, as we recall in Section \ref{relativeThreshold}, which motivates our choice.
Also, we cover the absolute threshold strategy considered in \cite{etienne2017periodic,etienne2017asynchronous,etienne2016event} as a special case.
 We show on an example that the addition of the scalar auxiliary variable can significantly reduce the number of transmissions compared to an absolute threshold rule, thereby providing a strong motivation for its use.
 
To analyze the proposed event-triggered observer, the overall plant-observer interconnection is modeled as a hybrid system using the formalism of \cite{goedel2012hybrid1, cai2009characterizations}, where a jump corresponds to an output transmission.  We show that the estimation error system satisfies a global practical stability property. 
The latter is not asymptotic in general mostly because we do not implement a copy of the observer in the triggering mechanism. 
 Moreover, the existence of a strictly positive minimum inter-event time is ensured under mild boundedness conditions on the plant state and its input. 
Finally, we apply the proposed approach in a numerical case study of a lithium-ion battery as mentioned above,
for which the number of transmissions can be significantly reduced compared to an absolute threshold strategy, while still ensuring good estimation performance. 

Various event-triggered observer-based control strategies are available in the literature, such as e.g., \cite{tanwani2015using, abdelrahim2017robust, dolk2016output, liu2015event}. Nevertheless, these do not cover event-triggered estimation as a particular case, as significant technical difficulties arise, in particular in ruling out Zeno phenomenon, when the plant state is not required to converge towards a given attractor.

The remainder of the paper is organized as follows. Preliminaries are reported in Section~\ref{Notation}. The model and the problem statement are presented in Section~\ref{ModelAndProblemStatement}. The proposed triggering rule is given in Section~\ref{TriggeringRuleAndHybridModel}, where we model the system as a hybrid system. In Section~\ref{Main result}, we analyze the obtained estimation error as well as the inter-event times. The numerical case study is reported in Section~\ref{ExampleBattery}. Finally, Section~\ref{Conclusions} concludes the paper. 

\section{Preliminaries}\label{Notation}
The notation $\R$ stands for the set of real numbers and $\Rlo:= [0, \infty)$ is the set of positive real numbers. We use $\Z$ to denote the set of integer numbers, $\Zo:= \{0,1,2,...\}$ and $\Zp:= \{1,2,...\}$. For a vector $x \in \R^n$, $|x|$ denotes its Euclidean norm. For a matrix $A \in \R^{n  \times m}, \norm{A}$ stands for its 2-induced norm. Given a real, symmetric matrix $P$, its maximum (minimum) eigenvalue is denoted as $\lambda_{\max}(P) \  (\lambda_{\min}(P))$.

 We consider hybrid systems in the formalism of \cite{goedel2012hybrid1}, \cite{cai2009characterizations}, namely
\begin{equation}
	\mathcal{H} \;:\; \left\{
	\begin{array}{rcll}
		\dot x &=& F(x,u), & \quad (x,u)\in \mathcal{C}, 
		\\
		x^+ &=& G(x,u),  &\quad (x,u)\in \mathcal{D},
	\end{array}
	\right.
\end{equation}
where 
$\mathcal{C}\subseteq \R^{n_x}$ is the flow set, 
$\mathcal{D}\subseteq \R^{n_x}$ is the jump set,
$F$ is the flow map and $G$ is the jump map. 
Solutions  
to system (1) are defined on  \textit{hybrid time domains}. A set 
$E\subset \Rlo\times \Zo$ is a \textit{compact 
	hybrid time domain} if $E = \bigcup_{j=0}^{J-1}([t_j, t_{j+1}], j)$
for some finite sequence of times $0=t_0\leq t_1\leq \ldots \leq t_{J}$
and it is a \textit{hybrid time domain} if for all $(T,J)\in E$, 
$E\cap ([0,T]\times \{0,1,\ldots, J\})$
is a compact hybrid time domain. Given a hybrid time domain $E$, we define $
\sup_{j}E := \sup \{j\in \Zo: \exists \, t\in \Rlo \textnormal{ such that } (t,j)\in E
\}$.
A \textit{hybrid signal} is a function defined on a hybrid time domain. A hybrid signal
$u : \dom \,u\, \to \R^{n_u}$ is called a \textit{hybrid input} if $u(\cdot, j)$ is measurable and locally essentially
bounded for each $j$.
A hybrid signal $x : \dom \,x \,\to \R^{n_x}$ is called a \textit{hybrid arc} 
if $x(\cdot, j)$
is locally absolutely continuous for each $j$. A hybrid arc $x : \dom \, x \, \to \R^{n_x}$ and a
hybrid input $u : \dom \, u\, \to \R^{n_u}$ form a \textit{solution pair} $(x,u)$ to $\mathcal{H}$ if 
$\dom \, x \, = \dom \, u$,
$(x(0, 0), u(0, 0)) \in \mathcal{C} \cup \mathcal{D}$, and
\begin{itemize}
	\item for all $j\in \Zo$ and almost all $t$ such that $(t,j)\in \dom \, x$, 
	$(x(t,j), u(t,j))\in \mathcal{C}$ and $\dot x = F(x(t,j), u(t,j))$;
	\item for all $(t,j)\in \dom \, x$ such that $(t,j+1)\in \dom  \,x$, 
	$(x(t,j),u(t,j))\in \mathcal{D}$ and $x(t,j+1) = G(x(t,j),u(t,j))$.
\end{itemize}

\section{Problem statement}\label{ModelAndProblemStatement}

\begin{figure*}
	\begin{center}
		\tikzstyle{blockB} = [draw, fill=blue!30, rectangle, 
		minimum height=2em, minimum width=3em]  
		\tikzstyle{blockG} = [draw, fill=MyGreen!40, rectangle, 
		minimum height=2em, minimum width=3em]
		\tikzstyle{blockR} = [draw, fill=red!40, rectangle, 
		minimum height=2em, minimum width=3em]
		\tikzstyle{blockO} = [draw,minimum height=1.5em, fill=orange!20, minimum width=4em]
		\tikzstyle{input} = [coordinate]
		\tikzstyle{blockW} = [draw,minimum height=1.5em, fill=white!20, minimum width=4em]
		\tikzstyle{input1} = [coordinate]
		\tikzstyle{blockCircle} = [draw, circle]
		\tikzstyle{sum} = [draw, circle, minimum size=.3cm]
		\tikzstyle{blockSensor} = [draw, fill=white!20, draw= blue!80, line width= 0.8mm, minimum height=10em, minimum width=13em]
		
		\begin{tikzpicture}[auto, node distance=2cm,>=latex , scale=0.98,transform shape] 
			
			\node [input, name=input] {};
			\node [blockW, above of=input, node distance=1.3cm] (plant) {  \footnotesize 
				$		\begin{array}{l}
					\dot{x} = Ax + Bu \\ y = Cx
				\end{array}$};

			\node [input, above of= plant, node distance=0.7cm, right = 0.5 cm] (plantName) {};
			\draw [] (plantName) -- node  [pos=0.5]{\textbf{Plant}} (plantName);
			
			\draw [draw,->] (input) -- node {} (plant);
			\draw [draw,-] (input) -- node [pos=0.5]{$u$} (plant);
			
			\node [blockW, right of= plant, node distance=3.5cm] (triggering) {\begin{minipage}{10em} \centering
					\footnotesize Transmit $\bar{y}$ when\\[.2em] 
					$\gamma |\bar{y} - y|^2 \geq \sigma c_1 \eta + \varepsilon$\\[.2em] 
					where\\[.2em] 
					$ \begin{array}{l}
						\dot{{\eta}} = -c_1\eta + c_2|\bar{y} - y|^2\\
						\eta^{+} = c_3\eta \\
						\dot{\bar{y}} =0, \;
						\bar{y}^+ = y		\end{array}$
			\end{minipage} };
			\node [input, above of= triggering, node distance=1.3cm, right = 1.1 cm] (triggeringName){};
			\draw [] (triggeringName) -- node  [pos=0.5]{\textbf{Smart sensor}} (triggeringName);
			\draw [draw,->] (plant) -- node  [pos=0.5]{$y$} (triggering);
			
			\node [cloud, draw,cloud puffs=10,cloud puff arc=120, aspect=2, inner ysep=0.7em, right of=triggering, node distance=3.5cm, 
			fill =white!30] (network){\textbf{Network}};
			\draw [draw,->] (triggering) -- node [pos=0.5]{} (network);
			
			\node [blockW, right of=network, node distance=3.7cm] (observer) {
				\footnotesize 
				$\begin{array}{l}
					\dot{\hat{x}} = A\hat{x} + Bu + L(\bar{y} - \hat{y})\\\hat{y} = C\hat{x}\end{array}	$
			};
			\node [input, above of= observer, node distance=0.7cm, right = 0.8 cm] (observerName){};
			\draw [] (observerName) -- node  [pos=0.5]{\textbf{Observer}} (observerName);
			\draw [draw,->] (network) -- node [pos=0.5]{$\bar{y}$} (observer);
			\node [input, below of=observer, node distance=1.3cm] (input2) {};
			\draw [draw,->] (input2) -- node [pos=0.5]{$u$} (observer);
			\node [input, right of=observer, node distance=2.7cm] (stateEstimate) {};
			\draw [draw,->] (observer) -- node [pos=0.5]{$\hat{x}$} (stateEstimate);
			\draw [draw,-] (input) -- (input2);
			\node [input, right of= input, node distance=5.37cm] (input3){};
			\node [input, below of= input3, node distance=0.4cm] (input4){};
			\draw [draw,-] (input3) -- (input4);
			
		\end{tikzpicture}
	\end{center}
	\caption{Block diagram representing the system architecture}
	\label{Fig:blockDiagram}
\end{figure*}
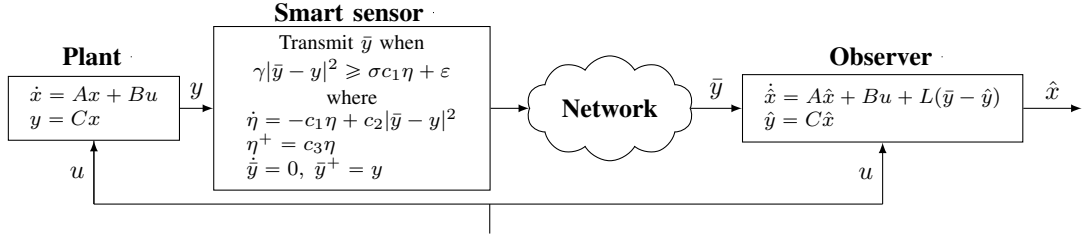

Consider the linear system
\begin{equation}
	\begin{aligned}
		\dot x  &=  Ax + Bu\\
		y  &=  Cx,
	\end{aligned}
	\label{eq:system}
\end{equation}
where $x \in \R^{n_x}$ is the state, $u \in \R^{n_u}$ is a known input, and $y \in \R^{n_y}$ is the measured output with $n_x$, $n_y$ $\in \Zp$ and $n_u \in \Zo$ .  
The pair $(A,C)$ is assumed to be detectable. Hence, by letting $L \in \R^{n_x  \times n_y}$ be any matrix such that $A-LC$ is Hurwitz, we can design a Luenberger observer \cite{luenberger1966observers} of the form 
\begin{equation}
	\begin{aligned}
		\dot{\hat x}  &=  A \hat x + Bu + L(y- \hat y)\\
		\hat y  &=  C\hat x,
	\end{aligned}
	\label{eq:observerAlly}
\end{equation}
where $\hat{x} \in \R^{n_x}$ is the state estimate.  
Observer \eqref{eq:observerAlly}, when it has access to input $u$ and measured output $y$ continuously, guarantees that we are able to asymptotically reconstruct the state $x$ of the plant, implying that $\displaystyle \lim_{t \rightarrow \infty}(x(t)-\hat{x}(t))=0$ for any initial condition to \eqref{eq:system} and \eqref{eq:observerAlly} and any input $u$.  In this work, we investigate the scenario where the plant measurement $y$ is transmitted to observer \eqref{eq:observerAlly} via a digital channel, see Fig.~\ref{Fig:blockDiagram}, and therefore only samples of $y$ are available to the observer. Moreover, since the output is sent via a packet-based network, we want to sporadically transmit it, while still achieving good estimation properties. Therefore, our goal is to design a triggering rule to decide when $y$ needs to be transmitted to observer \eqref{eq:observerAlly}, with the mentioned properties. We assume for this purpose that the sensor is ``smart" in the sense that it can run a local one-dimensional dynamical system. 
We also adopt the following assumption. 

\begin{ass}\itshape
	The observer has access to the input $u$ continuously. \hfill $\Box$ 
	\label{assumption 1}
\end{ass}
Assumption~\ref{assumption 1}  is a reasonable assumption in many control applications, such as, for example, when the control input is generated on the observer side. 
The relaxation of this assumption is left for future work. 
 
In this setting, the observer does not know $y$ but only its sampled version $\bar y$, which is generated with a zero-order-hold device between two successive transmission instants, i.e., in terms of the hybrid systems notation of Section~\ref{Notation}
\begin{equation}
	\dot{\bar y} = 0 \\
\end{equation}
and, when a transmission occurs the output is sampled, considering an ideal sampler, 
\begin{equation}
	\bar y^+ = y.
	\label{eq:yplus}
\end{equation}
The observer equations in \eqref{eq:observerAlly} are then modified to become
\begin{equation}
	\begin{aligned}
		& \dot{\hat x}  =  A \hat x + Bu + L(\bar y- \hat y)\\
		& \hat y  =  C\hat x.
		\label{eq:observerNew}
	\end{aligned}
\end{equation}
Defining the sampling-induced error $e:= \bar y - y $, we obtain
\begin{equation}
	\dot{\hat x}  =  A \hat x + Bu + L(y- \hat y + e).\\
	\label{eq:observer}
\end{equation}
The sampling-induced error $e$ dynamics between two successive transmission instants is 
\begin{equation}
	\dot{e}  = \dot{\bar{y}} - \dot{y} = -\dot{y} = -C\dot{x} = -CAx -CBu,
	\label{eq:samplingErrorContinuous}
\end{equation}
and, at each transmission instant we have $	e^+ = 0,$
in view of \eqref{eq:yplus}. Let $\xi := x -\hat x \in \R^{n_x}$ be the state estimation error. Its dynamics is, between two successive transmission instants, in view of \eqref{eq:system} and \eqref{eq:observer},
\begin{equation}
	\dot{\xi} =  (A -LC) \xi -Le
	\label{eq:estimationErrorContinuous}
\end{equation}
and, at each transmission instant, $\xi^+ = \xi.$ 

Our objective is to define a triggering rule, which ensures global practical asymptotic stability of estimation error dynamics and guarantees the existence of a positive minimum inter-event time between two consecutive transmissions.
\begin{rem}\itshape
	When the system output is of the form $y = Cx + Du +d$, where $d$ is a known constant, we can generate a new output $z = Cx$ by using the knowledge of $d$, the measured output $y$ and the input $u$, which is available thanks to Assumption \ref{assumption 1}.  The system then becomes of the form of \eqref{eq:system} again. We will exploit this observation in the example of Section~\ref{ExampleBattery}.
	\hfill $\Box$
	\label{Remark1}
\end{rem}

\section{Triggering rule and hybrid model}\label{TriggeringRuleAndHybridModel}
\subsection{Relative threshold is not suitable for estimation} 
\label{relativeThreshold}
We first note that the general event-triggered control solutions for stabilization may not be (directly) used for the estimation problem at hand. We illustrate this with the relative threshold technique developed for control in \cite{tabuada2007event} to define the triggering rule.
To see this, note that since $A-LC$ is Hurwitz, we can define $V: \xi \mapsto \xi^\top P\xi $ on $\R^{n_x}$, where $P\in \R^{n_x \times n_x}$ is symmetric, positive definite and verifies $(A-LC)^\top P + P(A-LC) = -Q$ for some  $Q \in \R^{n_x \times n_x}$ symmetric and positive definite. Then, for any $\xi \in \R^{n_{\xi}}$ and $ e \in \R^{n_y} $, 
\begin{equation}
	\left\langle \nabla V(\xi), (A-LC)\xi -Le \right\rangle \leq -\alpha V(\xi) +\gamma|e|^2,
	\label{eq:ISSKfunction}
\end{equation}
where $\displaystyle \alpha := \frac{\lambda_{\min}(Q)}{\lambda_{\max}(P)}(1-c) >0 $, $\displaystyle \gamma := \frac{\norm{PL}^2}{c\lambda_{\min}(Q)} >0 $ and $c \in (0,1)$ a design parameter.
We might then be tempted, in line with the design philosophy of \cite{tabuada2007event}, to define the triggering rule as
\begin{equation}
	\gamma|e|^2 \leq \varsigma\alpha V(\xi),	
	\label{eq:eventTriggering}
\end{equation}
with $\varsigma \in (0,1)$, which implies $ \left\langle \nabla V(\xi), (A-LC)\xi -Le \right\rangle \leq   - (1-\varsigma)\alpha V(\xi)$  and thus that $V$ strictly decreases along the solutions to \eqref{eq:estimationErrorContinuous}. 
However, \eqref{eq:eventTriggering} cannot be implemented because the estimation error $\xi$ is not available for the triggering rule, as it depends on $x$ and $\hat{x}$. 

\subsection{Dynamic triggering rule}
To overcome the issue presented in Section \ref{relativeThreshold}, we introduce a scalar auxiliary variable $\eta$, whose equations during flows and jumps are
\begin{equation}
	\begin{aligned}
	\dot{{\eta}} &= -c_1\eta + c_2|e|^2,\\
	\eta^+ &= c_3 \eta,
	\end{aligned}
	\label{eq: etaFlow}
\end{equation}
where $c_1 >0$, $c_2 \geq 0$ and $c_3 \in [0,1]$ are design parameters, that will be selected later according to Theorem \ref{thm1}.

\begin{rem} \itshape
	The choice of the dynamics \eqref{eq: etaFlow} is inspired by norm-estimators \cite{krichman2001input}. Indeed, if $c_1$ and $c_2$ are selected such that $c_1 = \alpha$ and $c_2 = \gamma$, $\eta$ in \eqref{eq: etaFlow} is a norm-estimator, according to \cite[Definition 2.4]{krichman2001input}, but this particular choice of $c_1$ and $c_2$ is not necessary for the proposed triggering rule.  \hfill $\Box$
\end{rem}

By collecting all the equations, we obtain the hybrid model
\begin{equation}
	\begin{split}
		\left.
		\begin{aligned}
			\dot{x} & = Ax + Bu \\
			\dot{\xi} & = (A -LC) \xi -Le\\
			\dot{e} & = -CAx -CBu \\
			\dot{\eta} & = -c_1\eta + c_2 |e|^2 \\
		\end{aligned}
		\right\rbrace
		\ \ \ \ \ (x, \xi, e, \eta,u)  \in \mathcal{C},\\
		\\
		\left.
		\begin{aligned}
			x^+ & = x\\
			\xi^+ & = \xi\\
			e^+ & = 0\\
			\eta^+ &= c_3 \eta\\
		\end{aligned}
		\right\rbrace
		\ \ \ \ \ (x, \xi, e, \eta,u)  \in \mathcal{D},\\
		\label{eq:HybridWithNormEstimator}
	\end{split}
\end{equation}
for which a jump corresponds to a transmission of the current value of $y$ to the observer. The triggering rule is implemented through the flow and jump sets, $\mathcal{C}$ and $\mathcal{D}$, which are defined as\footnote{We make the sets $\mathcal{C}$ and $\mathcal{D}$ depending on $(q,u)$, even if here it could be only $q \in \mathcal{C}$ and $q \in \mathcal{D}$. This choice is convenient for the analysis of the inter-event times that will be presented in Section \ref{IET}.}
\begin{equation}
	\mathcal{C} := \big\{(q,u) \in \R^{n_q}\times \R^{n_u}: \gamma|e|^2 \leq \sigma c_1 \eta + \varepsilon,  \eta \geq 0 \big\}
	\label{eq:FlowSet}
\end{equation}
\begin{equation}
	\mathcal{D} := \big\{(q,u) \in \R^{n_q}\times \R^{n_u}: \gamma|e|^2 \geq \sigma c_1 \eta + \varepsilon, \eta \geq 0 \big\},
	\label{eq:JumpSet}
\end{equation}
where $q$ is the overall state, defined as  $q := (x, \xi, e, \eta) \in \R^{n_q} = \R^{n_x} \times \R^{n_x} \times \R^{n_y} \times \R$, with $n_q := 2n_x + n_y+~1$.
Constant $\gamma$ in \eqref{eq:FlowSet}-\eqref{eq:JumpSet} comes from \eqref{eq:ISSKfunction}, $\sigma \geq 0 $ is a design parameter and $\varepsilon$ is a strictly positive constant needed to avoid the Zeno phenomenon. Indeed, we will prove in the sequel that there exists a minimum inter-event time between two consecutive jumps under mild extra conditions whenever $\varepsilon >0$. Sets $\mathcal{C}$ and $\mathcal{D}$ in \eqref{eq:FlowSet}-\eqref{eq:JumpSet} essentially mean that a transmission is triggered whenever $\gamma|e|^2 \geq \sigma c_1 \eta + \varepsilon $, see Fig. \ref{Fig:blockDiagram}. The condition that $\eta \geq 0$ in \eqref{eq:FlowSet}-\eqref{eq:JumpSet} never generates a transmission as it is always true whenever $\eta$ is initialized with a non-negative value. It is thus only specified in \eqref{eq:FlowSet}-\eqref{eq:JumpSet} to emphasize that $\eta$ only takes non-negative values.  
 It is worth noting that, when $\sigma = 0$, the triggering rule proposed in \eqref{eq:FlowSet}-\eqref{eq:JumpSet} corresponds to an absolute threshold triggering rule, as in, e.g., \cite{etienne2017periodic,etienne2017asynchronous,etienne2016event}. 
 
For the sake of convenience we write system \eqref{eq:HybridWithNormEstimator}-\eqref{eq:JumpSet} as
\begin{equation}
	\begin{aligned}
		\dot{q} &= f(q,u),&& (q,u) \in \mathcal{C}\\
		q^{+} &= g(q),   && (q,u) \in \mathcal{D}.
	\end{aligned}
	\label{eq:hybridSystemCompact}
\end{equation}
We are ready to proceed with the analysis of \eqref{eq:hybridSystemCompact}.

\section{Main result}\label{Main result}
\subsection{Stability} 
The next theorem explains how to select the design parameters $c_1$, $c_2$, $c_3$ and $\sigma$ in  \eqref{eq:hybridSystemCompact} in order to guarantee that the observer \eqref{eq:observerAlly} is able to globally practically estimate the state $x$ of system \eqref{eq:system} in the configuration explained in Section \ref{ModelAndProblemStatement}, in which the measured outputs are not available at all times but only when the triggering rule enables transmissions.  
\begin{thm}\label{thm1} \itshape
	Consider system \eqref{eq:hybridSystemCompact}, any $\bar{\alpha} \in (0, \alpha]$,  where $\alpha$ comes from \eqref{eq:ISSKfunction}, and any $\nu >0$, select $c_1$, $c_2$, $c_3$, $\sigma$ and $\varepsilon$ as follows. 
	\begin{enumerate}[label=(\roman*),leftmargin=.5cm]
		\item $c_2 \in [0, c_2^{*}]$ and $\sigma \in [0, \sigma^{*}] $, where $c_2^{*} \geq 0$ and $\sigma^{*} > 0$ are such that $\sigma^{*}c_2^{*} < \gamma$, where $\gamma$ comes from \eqref{eq:ISSKfunction}.
		\item $c_1 \geq c_1^{*}$, where $c_1^{*} > 0$ is such that $\displaystyle c_1^{*}>\bar{\alpha}\Big(1-\frac{\sigma^{*}c_2^{*}}{\gamma}\Big)^{-1}$.
		\item $c_3 \in [0,1]$.
		\item $\varepsilon \in (0, \varepsilon^{*}]$, where $\displaystyle \varepsilon^{*} = \nu\bar{\alpha}\gamma(\gamma + c_2^*d)^{-1}$ with $\displaystyle d:= \sigma^{*} \Big (1-\frac{\sigma^{*} c_2^{*}}{\gamma} -\frac{\bar{\alpha}}{c_1^{*}} \Big )^{-1} > 0$.
	\end{enumerate}	
	Then for any solution pair $(q, u)$ and any $(t,j) \in 
	\dom q$,
	\begin{equation}
		V(\xi(t,j)) +d\eta(t,j) \leq e^{-\bar{\alpha}t} (V(\xi(0,0)) + d\eta(0,0)) + \nu. 
		\label{LyapunovEquationTheorem}
	\end{equation}\hfill $\Box$	
\end{thm}
\noindent\textbf{Proof:}
Let all conditions of Theorem~\ref{thm1} hold. We consider the Lyapunov function candidate
\begin{equation}
	U(q) =  V(\xi) + d\eta,
	\label{eq: LyapunovU}
\end{equation}
for any $q \in \R^{n_q}$, where $\displaystyle d $ is defined in item (iv) of Theorem~\ref{thm1}; note that $d> 0$ in view of items (i) and (ii) of Theorem~\ref{thm1}. 

Let $(q,u) \in \mathcal{C}$, in view of \eqref{eq:ISSKfunction} and \eqref{eq:HybridWithNormEstimator}, 
\begin{equation}
\begin{array}{l}
		\left\langle \nabla U(q), f(q,u) \right\rangle  
\\[.5em]
\qquad = 		
		 \left\langle \nabla V(\xi), (A-LC)\xi -Le \right\rangle  + d( -c_1\eta + c_2|e|^2)\\[.5em]
		 \qquad \leq -\alpha V(\xi) + \gamma|e|^2 +d( -c_1\eta + c_2|e|^2) \\[.5em]
		\qquad = -\alpha V(\xi) - c_1d\eta + (\gamma + c_2d) |e|^2.
\end{array}
\end{equation}
Since $(q,u) \in \mathcal{C}$, we have $\gamma|e|^2 \leq \sigma c_1\eta + \varepsilon$, which is equivalent to $\displaystyle |e|^2 \leq \frac{\sigma c_1}{\gamma}\eta + \frac{\varepsilon}{\gamma}$ as $\gamma >0$. Hence, the next inequalities hold
\begin{equation}
\begin{array}{l}
		\left\langle \nabla U(q), f(q,u) \right\rangle 
		\\
		\quad \leq -\alpha V(\xi) - c_1d\eta + (\gamma + c_2d)\Big (\frac{\sigma c_1}{\gamma}\eta + \frac{\varepsilon}{\gamma} \Big) \\
		\quad = -\alpha V(\xi)  -c_1d\eta + (\gamma + c_2d)\frac{\sigma c_1}{\gamma}\eta + \frac{1}{\gamma}(\gamma + c_2d)\varepsilon\\
		\quad = -\alpha V(\xi) - c_1 \Big (1 -\frac{\sigma }{d}- \frac{\sigma }{\gamma}c_2 \Big )d\eta + \frac{1}{\gamma}(\gamma + c_2d) \varepsilon\\
		\quad\leq -\min \Big \{\alpha, c_1 \Big (1 -\frac{\sigma }{d}- \frac{\sigma }{\gamma}c_2 \Big ) \Big \} U(z)  + \frac{1}{\gamma}(\gamma + c_2d) \varepsilon.
	\end{array}\label{eq:LyapunovFlowParameters}
\end{equation}
Due to the choice of parameters $c_1$, $c_2$ and $\sigma$, we have that \eqref{eq:LyapunovFlowParameters} implies 
\begin{equation}
	\left\langle \nabla U(q), f(q,u) \right\rangle \leq -\bar{\alpha}U(q) + \frac{1}{\gamma}(\gamma + c_2d) \varepsilon.
	\label{eq:LyapunovFlowParameters2}
\end{equation}
Indeed, when $\min \Big \{\alpha, c_1 \Big (1 -\frac{\sigma }{d}- \frac{\sigma }{\gamma}c_2 \Big ) \Big \} = \alpha$, then $-\min \Big \{\alpha, c_1 \Big (1 -\frac{\sigma }{d}- \frac{\sigma }{\gamma}c_2 \Big ) \Big \} = -\alpha \leq -\bar{\alpha}$. Conversely, when $\min \Big \{\alpha, c_1 \Big (1 -\frac{\sigma }{d}- \frac{\sigma }{\gamma}c_2 \Big ) \Big \} = c_1 \Big (1 -\frac{\sigma }{d}- \frac{\sigma }{\gamma}c_2 \Big )$, which is strictly positive due to the definition of $d$ in item~(iv) of Theorem~\ref{thm1}, $\sigma$ and $c_2$,  we have
\begin{equation}
	\begin{aligned}
		-c_1 \Big (1 -\frac{\sigma }{d}- \frac{\sigma }{\gamma}c_2 \Big ) &\leq -c_1^{*}\Big (1 -\frac{\sigma }{d}- \frac{\sigma }{\gamma}c_2 \Big ) \\
		& \leq -c_1^{*}\Big (1 -\frac{\sigma^{*} }{d}- \frac{\sigma^{*} }{\gamma}c_2^{*} \Big )
	\end{aligned}
\end{equation}
and since $\displaystyle d= \sigma^{*} \Big (1-\frac{\sigma^{*} c_2^{*}}{\gamma} -\frac{\bar{\alpha}}{c_1^{*}} \Big )^{-1}$, we obtain
\begin{equation}
	-c_1 \Big (1 -\frac{\sigma }{d}- \frac{\sigma }{\gamma}c_2 \Big ) \leq -c_1^{*} \Big (1 -\frac{\sigma^{*} }{d}- \frac{\sigma^{*} }{\gamma}c_2^{*} \Big ) = -\bar{\alpha}.
\end{equation}
Hence, \eqref{eq:LyapunovFlowParameters2} holds and since $\displaystyle \varepsilon \leq \varepsilon^* = \nu\bar{\alpha}\gamma(\gamma + c_2^*d)^{-1}$ and $c_2 \leq c_2^*$,
\begin{equation}
	\begin{aligned}
		\left\langle \nabla U(q), f(q,u) \right\rangle &\leq -\bar{\alpha}U(q) + \frac{1}{\gamma}(\gamma + c_2d) \varepsilon\\
		& \leq -\bar{\alpha}U(q) + \frac{1}{\gamma}(\gamma + c_2^*d) \varepsilon^{*}\\
		& =  -\bar{\alpha}U(q) + \bar{\alpha}\nu.
	\end{aligned}
	\label{U_flows}
\end{equation}
Let $(q,u)$ in $\mathcal{D}$, in view of \eqref{eq:HybridWithNormEstimator} and since $c_3 \in [0,1]$,
\begin{equation}
	\begin{aligned}
		U(g(q)) &= V(\xi) + d c_3\eta \leq V(\xi) + d\eta = U(q).
	\end{aligned}
	\label{deltaU_jumps}
\end{equation}

We now follow similar steps as in \cite[proof of Theorem 3.18]{goedel2012hybrid1} to show that \eqref{LyapunovEquationTheorem} holds. 
Let $(q, u)$ be a solution pair to system \eqref{eq:hybridSystemCompact}. Pick any $(t,j) \in \dom q$ and let $0 = t_0 \leq t_1 \leq \dots \leq t_{j+1} = t$ satisfy $\dom \displaystyle q \cap ([0,t]\times \{0,1,\dots,j\}) = \bigcup_{i=0}^j [t_i, t_{i+1}] \times \{i\}$. For each $i \in \{0,\dots,j\}$ and almost all $s \in [t_i, t_{i+1}]$, $(q(s,i), u(s,i)) \in \mathcal{C}$. Then, \eqref{U_flows} implies that, for each $i \in \{0,\dots, j\}$ and for almost all $s \in [t_i, t_{i+1}]$, 
\begin{equation}
	\frac{d}{ds}U(q(s,i)) \leq -\bar{\alpha}U(q(s,i)) + \bar{\alpha}\nu. 
\end{equation}	
Applying the comparison principle \cite[Lemma 3.4]{khalil2002nonlinear}, we obtain, for all $(s, i) \in  \dom q$,
\begin{equation}
	\begin{aligned}
		U(q(s,i)) & \leq e^{-\bar{\alpha}(s-t_i)}U(q(t_i,i)) + \bar{\alpha}\nu \int_{t_i}^{s} e^{-\bar{\alpha}(s-\tau)} d\tau\\
		& = e^{-\bar{\alpha}(s-t_i)}U(q(t_i,i)) + \bar{\alpha}\nu \frac{1}{\bar{\alpha}}[1-e^{-\bar{\alpha}(s-t_i)}].\\
	\end{aligned}
\label{eq:ULyapunovComparisonPrinciple}
\end{equation}
Thus, 
\begin{equation}
	U(q(t_{i+1},i))\leq e^{-\bar{\alpha}(t_{i+1}-t_i)}U(q(t_i,i)) + \nu - \nu e^{-\bar{\alpha}(t_{i+1}-t_i)} 
	\label{eq: solutionFlows}
\end{equation}
for all  $i \in \{0,\dots,j\}.$ 
Similarly, for each $i\in \{1,\dots,j\}$, $q(t_i,i-1) \in \mathcal{D}$. From \eqref{deltaU_jumps}, we obtain
\begin{equation}
	U(q(t_i,i)) - U(q(t_i,i-1)) \leq 0 \ \ \ \forall i \in \{1,\dots,j\}.
	\label{eq: solutionsJumps}
\end{equation}
From \eqref{eq:ULyapunovComparisonPrinciple}, \eqref{eq: solutionFlows} and \eqref{eq: solutionsJumps}, we can deduce that for any $(t,j) \in \dom q$,
\begin{equation}
	\begin{aligned}
	U(q(t,j)) & \leq e^{-\bar{\alpha}t}U(q(0,0)) + \nu - \nu e^{-\bar{\alpha}t}\\
	&\leq e^{-\bar{\alpha}t}U(q(0,0)) + \nu.
	\end{aligned}
\end{equation}
On the other hand, from \eqref{eq: LyapunovU}, we have
\begin{equation}
	\begin{aligned}
		 U(q(t,j)) &\leq  e^{-\bar{\alpha}t}U(q(0,0)) + \nu\\
		&= e^{-\bar{\alpha}t}[V(\xi(0,0)) + d\eta(0,0))] + \nu,
	\end{aligned}
\end{equation}
which concludes the proof as $U(q(t,j)) = V(\xi(t,j)) +d\eta(t,j)$. \hfill $\blacksquare$

It is important to note that, in absence of a digital network between the plant and the observer (i.e., when $e = 0$), we have from \eqref{eq:ISSKfunction} that for any solution $\xi$ to $\dot{\xi} = (A-LC)\xi$,  $V(\xi(t)) \leq e^{-\alpha t} V(\xi(0))$ for all $t \geq 0$. 
In view of \eqref{LyapunovEquationTheorem}, and as $d>0$, for any solution pair $(q,u)$ to \eqref{eq:hybridSystemCompact}, since $\eta$ takes non-negative values in view of \eqref{eq:FlowSet}-\eqref{eq:JumpSet}, $V(\xi(t,j)) \leq e^{-\bar{\alpha}t} (V(\xi(0,0)) + d\eta(0,0)) + \nu$.
 Hence, we guarantee a convergence rate $\bar{\alpha} \in (0,\alpha]$ of $V$ along the $\xi$-component of the solution to \eqref{eq:hybridSystemCompact}, which can be equal to $\alpha$. 
 We also have $\nu$ in \eqref{LyapunovEquationTheorem}, which is an ultimate bound of the estimation error, that is tuneable and can thus be made arbitrarily small (by selecting $\varepsilon$ small mainly) irrespective of the chosen convergence rate at the price of more frequent transmissions in general. Property \eqref{LyapunovEquationTheorem} also ensures that the auxiliary variable $\eta$ is bounded and converges to a neighborhood of $0$.
 
 	 In Theorem~\ref{thm1}, we first fix a convergence rate $\bar{\alpha}$ and a guaranteed ultimate bound $\nu$ for $V(\xi) + d\eta$, and then we explain how to select the design parameters to accomplish this. 
 	 It is worth noting that the conditions of Theorem~\ref{thm1} can be always ensured. Indeed, we just have to select $\sigma^*$ and $c_2^*$ sufficiently small such that $\sigma^* c_2^* < \gamma$, which is always possible, and all the other parameters can be always selected such that items (ii)-(iv) of Theorem~\ref{thm1} are verified as well. 
 Another way to use the result of Theorem~\ref{thm1} is to select  $\sigma$ and $c_2$ such that $\sigma c_2 < \gamma$ holds. Then, by selecting $c_3 \in [0,1]$ and any strictly positive value for $c_1$ and $\varepsilon$, \eqref{LyapunovEquationTheorem} holds for some strictly positive $\bar{\alpha}$ and $\nu$. This is how we select parameters in the example in Section~\ref{ExampleBattery}.  
\subsection{Properties of the Inter-Event Times}\label{IET}
In this section we provide properties of the inter-event times. In particular, we first show the existence of a strictly positive minimum inter-event time between two consecutive transmissions under mild boundedness conditions on plant \eqref{eq:system}. This corresponds to the existence of a dwell-time for the solutions to \eqref{eq:hybridSystemCompact}, as defined in \cite{goedel2012hybrid1}, see, e.g., \cite{borgers2014event}, \cite{postoyan2014framework}.
 From the definitions of $\mathcal{C}$ and $\mathcal{D}$ in \eqref{eq:FlowSet} and \eqref{eq:JumpSet}, the inter-event time is lower bounded by the time that it takes for $|e|^2$ to grow from $0$, that is the $e$ value after a jump according to \eqref{eq:HybridWithNormEstimator}, to $\displaystyle \frac{\varepsilon}{\gamma}$. 
A proof that this time is bounded from below by a positive constant can be obtained by establishing that the time-derivative of $|e|^2$ is bounded. 
For this purpose, recalling that, from \eqref{eq:HybridWithNormEstimator} we have  $\dot{e} = -CAx -CBu$, we define the following set
\begin{equation}
	\mathcal{S}_{M} = \{(q,u) \in \R^{n_q} \times \R^{n_u}: |CAx + CBu| \leq M\},
	\label{eq: Set_boundedStateAndInput}
\end{equation}
where $M$ is an arbitrarily large positive constant. We restrict the flow and the jump sets  of system \eqref{eq:hybridSystemCompact} so that 
\begin{equation}
	\begin{aligned}
		\dot{q} &= f(q,u),&& (q,u) \in \mathcal{C}'\; := \mathcal{C} \cap \mathcal{S}_{M}\\
		q^{+} &= g(q),   && (q,u) \in \mathcal{D}':= \mathcal{D} \cap \mathcal{S}_{M}.
	\end{aligned}
	\label{eq:hybridSystemCompactNew}
\end{equation}
By doing so, we therefore only consider solutions to \eqref{eq:hybridSystemCompact} such that the derivative of $e$ is bounded. Hence, \eqref{LyapunovEquationTheorem} still applies. 
Note that 
\eqref{eq: Set_boundedStateAndInput} is verified for all hybrid times when the state $x$ and the input $u$ are known to lie in a compact set for all positive times and the constant $M$ is selected sufficiently large for instance. It is important to notice that the constraint \eqref{eq: Set_boundedStateAndInput} does not need to be implemented in the triggering rule: it is only used here for analysis purposes. 

In the next theorem we prove that there exists a positive minimum inter-event time between any two consecutive transmissions for solutions to system \eqref{eq:hybridSystemCompactNew}.  

\begin{thm} \itshape
	Consider system \eqref{eq:hybridSystemCompactNew}, then any solution pair $(q,u)$ has a dwell-time $\tau:= \displaystyle \frac{1}{2M} \sqrt{\frac{\varepsilon}{\gamma}}$ , i.e., for any $(s,i), (t,j) \in \dom q$ with $s+i \leq t+j$, we have $\displaystyle j-i \leq \frac{t-s}{\tau} +1$. 
	\hfill $\Box$
	\label{thm2}
\end{thm}

\noindent\textbf{Proof:}
Let $(q, u)$ be a solution pair to system \eqref{eq:hybridSystemCompactNew}. Pick any $(t,j) \in \dom q$ and let $0 = t_0 \leq t_1 \leq \dots \leq t_{j+1} = t$ satisfy $\dom \displaystyle q \cap ([0,t]\times \{0,1,\dots,j\}) = \bigcup_{i=0}^j [t_i, t_{i+1}] \times \{i\}$. For each $i \in \{0,\dots,j\}$ and almost all $s \in [t_i, t_{i+1}]$, $(q(s,i), u(s,i))~\in~\mathcal{C'}$. Then, from \eqref{eq:HybridWithNormEstimator} 
for all $s \in [t_i,t_{i+1}]$,
\begin{equation}
	\begin{array}{l}\displaystyle
		\frac{d}{ds}|e|^2 = \frac{d}{ds}(e^\top e) = (\dot{e}^\top e + e^\top \dot{e})\\
		\quad = (-CAx -CBu)^\top e + e^\top (-CAx -CBu)\\
		\quad = -2e^\top (CAx +CBu)\\
		\quad \leq \displaystyle 2|e||CAx + CBu|.
	\end{array}
\end{equation}
Since $(q(s,i), u(s,i)) \in \mathcal{C}' = \mathcal{C} \cap \mathcal{S}_{M}$, in view of \eqref{eq: Set_boundedStateAndInput},
\begin{equation}
	\frac{d}{ds}|e|^2 \leq 2|e|M.
	\label{eq:derivative_e2}
\end{equation}
Let $t_i' := \inf  \Big \{t \geq t_i:|e(t,i)| = \displaystyle \sqrt{\frac{\varepsilon}{\gamma}}  \Big \}$, hence $t_i' \leq t_{i+1}$ in view of \eqref{eq:JumpSet}.
For almost all $s \in [t_i,t_i']$, from \eqref{eq:derivative_e2}, we have
\begin{equation}
	\frac{d}{ds}|e|^2 \leq 2\sqrt{\frac{\varepsilon}{\gamma}}M.
\end{equation}
Integrating this equation and applying the comparison principle \cite[Lemma 3.4]{khalil2002nonlinear}, we obtain, for all $s \in [t_i,t_{i}']$
\begin{equation}
	|e(s,i)|^2 \leq |e(t_i, i)|^2 + 2\sqrt{\frac{\varepsilon}{\gamma}}M(s-t_i). 
\end{equation}
Moreover, since $e(t_i,i) = 0$, we obtain
\begin{equation}
	|e(s,i)|^2 \leq 2\sqrt{\frac{\varepsilon}{\gamma}}M(s-t_i) \ \ \forall s \in [t_i, t_{i}']. 
	\label{eq:SamplingErrorBounded}
\end{equation}
In view of \eqref{eq:SamplingErrorBounded}, $\displaystyle s \mapsto 2\sqrt{\frac{\varepsilon}{\gamma}}M(s-t_i)$ upper bounds $s\mapsto|e(s,i)|^2$ on  $[t_i, t_{i}']$. Hence, the time it takes for $\displaystyle s\mapsto2\sqrt{\frac{\varepsilon}{\gamma}}M(s-t_i)$  to grow from $0$ to $\displaystyle \frac{\varepsilon}{\gamma}$ is a lower bound on $t_{i}'- t_i \leq t_{i+1}- t_i $. Therefore, the solution $(q,u)$ has a dwell-time $\displaystyle \tau = \frac{1}{2M}\sqrt{\frac{\varepsilon}{\gamma}}$.  \hfill $\blacksquare$

 From Theorem \ref{thm2}, we see that the guaranteed minimum inter-event time $\tau$ grows when $M$ decreases or when $\varepsilon$ increases, which corresponds to an increase of the ultimate bound $\nu$, as shown in Theorem~\ref{thm1}.
Note that, because of \eqref{LyapunovEquationTheorem}, the $\eta$ and the $\xi$ components of the solutions to system \eqref{eq:hybridSystemCompactNew} cannot blow up in finite continuous time. In addition, if the constraint on the state $x$ and the input $u$ in \eqref{eq: Set_boundedStateAndInput} is satisfied for all continuous time $t \geq 0$, then we can ensure the 
 $t$-completeness of maximal solutions to system \eqref{eq:hybridSystemCompactNew}, see \cite[Definition 2.5]{goedel2012hybrid1}. As the conditions on $x$ and $u$ are assumptions on the original system \eqref{eq:system}, and not part of our design, we can indeed establish that $t$-completeness of maximal solutions to  \eqref{eq:hybridSystemCompactNew} is guaranteed, under appropriate assumptions on the initial states of $\eta$ and $\xi$, and thus a positive lower bounded on the inter-event times is guaranteed. Although this already sketches the main arguments, a complete and formal proof  will be given in future work.
 
 An additional feature of the proposed triggering rule is that it stops transmitting when the sampling-induced error $|e|$ becomes small enough, as formalized in the next lemma. 
 \begin{lem}
 	Consider system \eqref{eq:hybridSystemCompact}, given a solution pair $(q,u)$, if there exists $(t,j) \in \dom q$ such that $|e(t',j')| < \displaystyle \sqrt{\frac{\varepsilon}{\gamma}}$ for all $(t',j') \in \dom q$ with $t' + j' \geq t+j$, then 
 	$ \sup_j \dom q = j'< \infty $.
 	\hfill $\Box$
 	\label{lemma1}
 \end{lem} 
\noindent\textbf{Proof:} The condition $|e(t',j')| < \displaystyle \sqrt{\frac{\varepsilon}{\gamma}}$ for all $(t',j') \in \dom q$ with $t' + j' \geq t+j$ implies $\displaystyle \gamma |e(t',j')|^2 < \gamma \frac{\varepsilon}{\gamma} \leq \sigma c_1 \eta + \varepsilon$ for all $(t',j') \geq (t,j)$. Thus, the triggering condition is never triggered after $(t,j)$, hence no jumps occur after $(t,j)$ and $j' = j$ consequently. Therefore $ \sup_j \dom q = j'< \infty $, which concludes the proof.  \hfill $\blacksquare$

The condition on $|e|$ in Lemma~\ref{lemma1} occurs when the plant output $y$ remains for all positive times in a small neighborhood of a constant for instance. 
Indeed, when the output to plant \eqref{eq:system} satisfies $\displaystyle |y(t) - y^*| < \frac{1}{2}\sqrt{\frac{\varepsilon}{\gamma}}$ for all $t \geq T$ for some $T \geq 0$ and some constant $y^*\in \R^{n_y}$, we have for any solution pair $(q,u)$ to system \eqref{eq:hybridSystemCompactNew}, for any $(t_j,j), (t,j) \in \dom q$ with $(t_j,j-1) \in \dom q$ and $t_j \geq T$, $t\geq t_j$ and $\displaystyle |e(t,j)| = |y(t_j,j) - y(t,j)| = |y(t_j,j) - y^* + y^* - y(t,j)| \leq |y(t_j,j) - y^*|+ |y^* - y(t,j)| < 2 \frac{1}{2}\sqrt{\frac{\varepsilon}{\gamma}}$ and the condition of Lemma~\ref{lemma1} holds.
Moreover, it automatically starts transmitting again if that condition is no longer verified. This is a clear advantage over time-triggered strategies, where the measured output is always transmitted, which may be important in practical applications.
The above condition of $y$ of Lemma~\ref{lemma1} is verified, for example, when the plant is asymptotically stable and the input $u$ is constant, see also the example in the next section. 
Note that Lemma~\ref{lemma1} applies to system \eqref{eq:hybridSystemCompact}, and not only to system \eqref{eq:hybridSystemCompactNew}.
\section{Numerical case study }	\label{ExampleBattery}
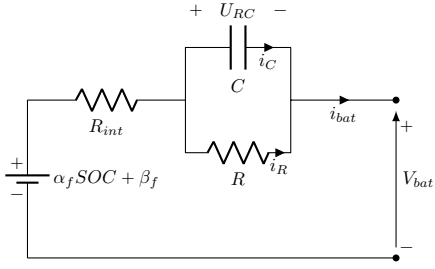
\begin{figure}[tbp]
	\begin{center}
		\begin{circuitikz}[
			scale=0.7,transform shape]\draw
			(0,0) to[R,l_=$R$,i_=$i_{R}$](2,0)--(2,2)
			(0,2) to[C,l_=$C$,i_=$i_{C}$](2,2) 
			(0,0) -- (0,2)
			(2,1) to [short,i_=$i_{bat}$,-*] (4,1)
			(-3,1) to [R,l_=$R_{int}$](0,1)
			(-3,1) to [battery1](-3,-2)
			node at (-3.2,-0.2) {$+$}
			node at (-3.2,-0.8) {$-$}
			(-3,-2) to [short,-*] (4,-2)
			node at (4.2,-1.8) {$-$}
			node at (4.2,0.5) {$+$}
			node at (0.2,2.7) {$+$}
			node at (1.8,2.7) {$-$}
			node at (1,2.7) {$U_{RC}$}
			node at (-1.5,-0.5) {$\alpha_fSOC + \beta_f$}
			[-latex] (4,-1.8) -> (4,0.8) node[midway,right] {$V_{bat}$};
			;
		\end{circuitikz}
	\end{center}
	\caption{Equivalent electrical circuit of a single battery cell}
	\label{FIG:batteryCircuit}
\end{figure}

We apply the proposed event-triggered observer to a lithium-ion battery example  \cite{he2012comparison}. This can be relevant when the battery management system is not co-located with the battery and communicates with it via a digital network. 
The considered electrical equivalent circuit of the battery cell 
 is shown in Fig.~\ref{FIG:batteryCircuit}. From the circuit, the following system model is derived
\begin{equation}
	\begin{aligned}
		\dot{U}_{RC} &= -\frac{1}{\tau} U_{RC} + \frac{1}{C}i_{bat}\\
		\dot{SOC} &= -\frac{1}{Q} i_{bat}\\
		V_{bat} &= -U_{RC} + \alpha_fSOC + \beta_f -R_{int}i_{bat}.
	\end{aligned}
	\label{eq: batteryModel}
\end{equation}
The states $U_{RC} \in \R$ and $SOC \in \R$ are the voltage on the $RC$ circuit and the battery state of charge, respectively. The input $i_{bat} \in \R$ is the battery current and the output $V_{bat} \in \R$ is the battery voltage. Considering the temperature to be constant and equal to $25\,^\circ$C, the following values are taken $\tau = 7$~s, $ C = 2.33 \cdot 10^4$~F, $Q = 25$~Ah, $R_{int}~=~4$~m$\Omega$, $ \alpha_f = 0.6$ and $\beta_f = 3.4$, which have been derived from experimental data. 
We design observer \eqref{eq:observerAlly} with $L = [0.64, 2.33]$. As a result, \eqref{eq:ISSKfunction} holds with $P = \begin{bmatrix}
	1.57 \cdot 10^4 & -3.39 \cdot 10^3\\
	-3.39 \cdot 10^3 & 1.29 \cdot 10^3
\end{bmatrix}$, $Q = \begin{bmatrix}
100 & 0 \\
0 & 1000
\end{bmatrix}$, 
$\alpha = 0.003$ and $\gamma = 1.104 \cdot 10^5$. 

From \eqref{eq: batteryModel}, we see that the system output has a feedthrough term, indeed, the output equation has the following structure $y = Cx + Du + \beta_f$. However, since the observer has access to the input $u = i_{bat}$ continuously thanks to Assumption~\ref{assumption 1} and $\beta_f$ is known, we can rewrite the output equation as $z = Cx$, as explained in Remark~\ref{Remark1}. 

\begin{figure}
	\centering
	\includegraphics[trim= 4.3cm 7.5cm 4cm 7.5cm, clip, width=1\linewidth]{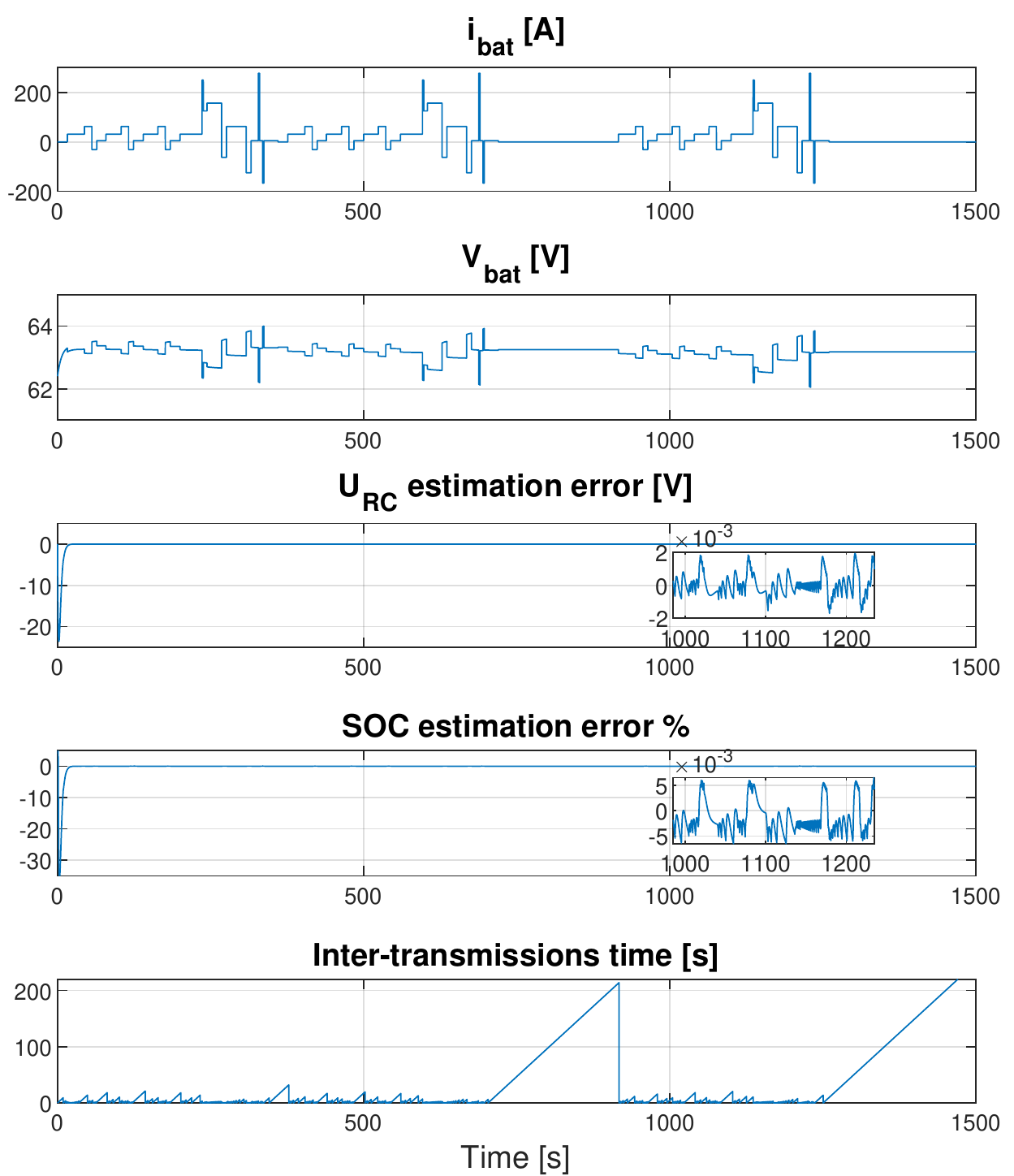}
	\caption{Input $i_{bat}$, output $V_{bat}$, state estimation error $\xi_{U_{RC}}$ and $\xi_{SOC}$, and inter-transmissions time, with $\sigma =500$, $c_1 = 1$, $c_2 = 50$, $c_3 = 1$, $\varepsilon = 1$.}
	\label{fig:input_output_IET}
\end{figure}

We have first simulated the event-triggered observer with $\sigma = 500$, $c_1 = 1$, $c_2 = 50$, $c_3 = 1$, $\varepsilon = 1$. With this choice of parameters, the condition $\sigma c_2 < \gamma$ is satisfied. 
The input is given by a plug-in hybrid electric vehicle (PHEV) current profile, shown in Fig.~\ref{fig:input_output_IET}, for which the solutions to \eqref{eq: batteryModel} remains in a compact set, so that $|CAx + CBu| \leq M$ for $M$ large enough along the solutions like in \eqref{eq: Set_boundedStateAndInput} and Theorem~\ref{thm2} applies. Fig.~\ref{fig:input_output_IET} also provides the plots of the corresponding output, state estimation error and inter-transmission times obtained with the following initial conditions: $U_{RC} (0,0) = 1$~V, $SOC (0,0) = 100$\%, $\xi_{U_{RC}} (0,0) = 0$~V, $\xi_{SOC} (0,0) = 75 $\%, $e(0,0) = 0$ and $\eta(0,0) = 10^6$. The minimum-inter event time seen in simulation is $0.227 \,$s. It is clear that both state estimation errors practically converge to zero. Moreover, the proposed scheme stops the transmissions whenever voltage $V_{bat}$ tends to a constant, like in  $[720 \,\textnormal{s},900 \, \textnormal{s}]$ and $[1260 \, \textnormal{s},1500 \,\textnormal{s}]$, where the inter-transmission time keeps growing, which is again a clear advantage over time-triggered policies. 
 Indeed, when the input $i_{bat} = 0$, the output $V_{bat}$ tends to constant and no data are transmitted, as explained in Lemma~\ref{lemma1}.
  Moreover, the transmissions start again when the input becomes different from 0. 

We have also analyzed the impact of the design parameters, in particular we focus on the effect of $\sigma$, $c_1$ and $\varepsilon$. For this purpose, we have simulated the corresponding system \eqref{eq:hybridSystemCompact} with different parameters configurations and $100$ different initial conditions each time, which were selected randomly in the interval $(0,3) V$ for $U_{RC} (0,0)$ and  $\xi_{U_{RC}} (0,0)$ and in the interval $(0,100)$\% for $SOC (0,0)$ and $\xi_{SOC} (0,0)$. The scalar variable $\eta$ and the sampling induced error were always initialized as $\eta(0,0) = 10^6$ and $e(0,0) = 0$. 
  For each choice of parameters, we have evaluated how many transmissions occur in the time interval $[0 \,\textnormal{s},1500 \, \textnormal{s}]$ on average as well the maximum absolute value of the state estimation errors $|\xi_{U_{RC}}(t,j)|$ and $|\xi_{SOC}(t,j)|$ with $t \in [1000 \, \textnormal{s}, 1500 \, \textnormal{s}]$ averaged over all simulations. 
The data collected are shown in Table~\ref{table_example}.

\begin{table}[t]
	\caption{\normalfont{Average number of transmissions in the time interval $[0 \,\textnormal{s},1500 \,\textnormal{s}]$
		, maximum absolute value of the state estimation errors $|\xi_{U_{RC}}(t,j)|$ and $|\xi_{SOC}(t,j)|$ for $t \in [1000 \,\textnormal{s},1500\,\textnormal{s}] $ 
		with different choices for $\sigma$, $c_1$, $\varepsilon$.}}
	\label{table_example}
	\begin{center}
		\begin{tabular}{ccc|ccc}
			 \toprule
			$\sigma$& $c_1$& $\varepsilon$ & Transmissions & $|\xi_{U_{RC}}|$ [V] & $|\xi_{SOC}|$\% \\
			\toprule
			$500$&$1$& $1$& $390 $  &$ 0.0019$&$ 0.0074 $\\
			\midrule
			$500$&$1$& $0.1$ & $1301 $ &$ 0.0006$&$ 0.0025 $\\
			$500$&$1$& $10$ & $ 102$  &$ 0.0067$&$ 0.0251 $\\
			$500$&$1$& $100$ & $19 $  &$ 0.0163$&$ 0.0754 $\\
			\midrule
			$500$&$0.01$& $1$ & $10 $  &$ 0.0171$&$ 0.0653 $\\
			$500$&$0.1$& $1$ & $340 $  &$ 0.0019$&$ 0.0069 $\\
			$500$&$10$& $1$ & $681 $  &$0.0021 $&$ 0.0077 $\\
			\midrule
			$1000$&$1$& $1$ & $364 $ &$ 0.0021$&$ 0.0082 $\\
		    $0$&$1$& $1$ & $886 $  &$ 0.0018$&$ 0.0069 $\\
		    \bottomrule
		\end{tabular}
	\end{center}
\end{table}

Table~\ref{table_example} shows that, in all considered configurations, the estimation error is small. Moreover, the data suggest that there is a trade-off between the number of transmissions and the estimation accuracy, as already indicated in Section~\ref{Main result}. In particular, when $\varepsilon$ is small, we have more transmissions, but the error is smaller. Conversely, when $\varepsilon$ is large, the number of transmissions is reduced, but the estimation error increases, even if it is still reasonably small in view of the application. Moreover, Table~\ref{table_example} shows that the larger $c_1$, the higher the number of transmissions required, without a big impact on the accuracy of the estimation error, except from the case when $c_1 = 0.01$ which produces only $10$ transmissions, but the estimation error is higher. 
Furthermore, there is a trade-off also on the choice of $\sigma$. Indeed, the larger $\sigma$, the smaller the number of transmissions, but the larger the error. It is important to note that the last parameters choice in Table~\ref{table_example}, with $\sigma = 0$, corresponds to an absolute threshold triggering rule and leads to many transmissions. 

\section{Conclusions} \label{Conclusions}
We have presented an event-triggered observer design for linear time-invariant systems. In order to reduce the number of transmissions over a network while still ensuring good estimation performance, we have proposed a dynamic triggering rule, implemented by a smart sensor, which decides when the measured output needs to be transmitted to the observer. Compared with other works in the literature, we do not need a copy of the observer in the sensor, but only a first order filter of the sampling-induced error, which may allow to significantly reduce the number of transmissions compared to an absolute threshold policy, while being easily implementable. 

 We have modeled the system as a hybrid system and we have shown that the estimation error system satisfies a global practical stability property. Moreover, under mild boundeness conditions on the plant state and its input, we have proved that the system does not exhibit the Zeno phenomenon and even has a positive lower bound on the inter-event times. 
 
In future work, we plan to extend the results to nonlinear systems assuming  the estimation error system satisfies an input-to-state stability property, see, for instance, \cite{astolfi2021stubborn}.
 We will also include measurement noise and disturbances in the system model. Moreover, we will investigate the relaxation of Assumption~\ref{assumption 1} and prove that maximal solutions are complete.
 
\bibliography{bibliography}

\begin{thebibliography}{10}

\bibitem{postoyan2011framework}
R.~Postoyan and D.~Ne{\v{s}}i{\'c}, ``A framework for the observer design for
  networked control systems,'' {\em IEEE Transactions on Automatic Control},
  vol.~57, no.~5, pp.~1309--1314, 2011.

\bibitem{li2017robust}
Y.~Li, S.~Phillips, and R.~G. Sanfelice, ``Robust distributed estimation for
  linear systems under intermittent information,'' {\em IEEE Transactions on
  Automatic Control}, vol.~63, no.~4, pp.~973--988, 2017.

\bibitem{ferrante2016state}
F.~Ferrante, F.~Gouaisbaut, R.~G. Sanfelice, and S.~Tarbouriech, ``State
  estimation of linear systems in the presence of sporadic measurements,'' {\em
  Automatica}, vol.~73, pp.~101--109, 2016.

\bibitem{mazenc2015design}
F.~Mazenc, V.~Andrieu, and M.~Malisoff, ``Design of continuous--discrete
  observers for time-varying nonlinear systems,'' {\em Automatica}, vol.~57,
  pp.~135--144, 2015.

\bibitem{li2010event}
L.~Li, M.~Lemmon, and X.~Wang, ``Event-triggered state estimation in vector
  linear processes,'' in {\em Proceedings of the American control conference,
  \textnormal{Baltimore, MD, USA}}, pp.~2138--2143, 2010.

\bibitem{shi2014event2}
D.~Shi, T.~Chen, and L.~Shi, ``Event-triggered maximum likelihood state
  estimation,'' {\em Automatica}, vol.~50, no.~1, pp.~247--254, 2014.

\bibitem{li2011performance}
L.~Li and M.~Lemmon, ``Performance and average sampling period of sub-optimal
  triggering event in event triggered state estimation,'' in {\em 50th IEEE
  Conference on Decision and Control and European Control Conference,
  \textnormal{Orlando, FL, USA}}, pp.~1656--1661, 2011.

\bibitem{trimpe2014stability}
S.~Trimpe, ``Stability analysis of distributed event-based state estimation,''
  in {\em 53rd IEEE Conference on Decision and Control, \textnormal{Florence,
  Italy}}, pp.~2013--2019, 2014.

\bibitem{scheres2021Event}
K.~J.~A. Scheres, M.~S.~T. Chong, R.~Postoyan, and W.~P. M.~H. Heemels,
  ``Event-triggered state estimation with measurement noise,'' {\em 60th IEEE
  Conference on Decision and Control, \textnormal{Austin, TX, USA}}, 2021.

\bibitem{andrieu2015self}
V.~Andrieu, M.~Nadri, U.~Serres, and J.-C. Vivalda, ``Self-triggered
  continuous--discrete observer with updated sampling period,'' {\em
  Automatica}, vol.~62, pp.~106--113, 2015.

\bibitem{rabehi2020finite}
D.~Rabehi, N.~Meslem, and N.~Ramdani, ``Finite-gain event-triggered interval
  observers design for continuous-time linear systems,'' {\em International
  Journal of Robust and Nonlinear Control}, vol.~31, no.~9, pp.~4131--4153,
  2021.

\bibitem{han2015stochastic}
D.~Han, Y.~Mo, J.~Wu, S.~Weerakkody, B.~Sinopoli, and L.~Shi, ``Stochastic
  event-triggered sensor schedule for remote state estimation,'' {\em IEEE
  Transactions on Automatic Control}, vol.~60, no.~10, pp.~2661--2675, 2015.

\bibitem{shi2016event}
D.~Shi, T.~Chen, and M.~Darouach, ``Event-based state estimation of linear
  dynamic systems with unknown exogenous inputs,'' {\em Automatica}, vol.~69,
  pp.~275--288, 2016.

\bibitem{huang2019robust}
J.~Huang, D.~Shi, and T.~Chen, ``Robust event-triggered state estimation: A
  risk-sensitive approach,'' {\em Automatica}, vol.~99, pp.~253--265, 2019.

\bibitem{etienne2017periodic}
L.~Etienne, S.~Di~Gennaro, and J.-P. Barbot, ``Periodic event-triggered
  observation and control for nonlinear {L}ipschitz systems using impulsive
  observers,'' {\em International Journal of Robust and Nonlinear Control},
  vol.~27, no.~18, pp.~4363--4380, 2017.

\bibitem{etienne2017asynchronous}
L.~Etienne, Y.~Khaled, S.~Di~Gennaro, and J.-P. Barbot, ``Asynchronous
  event-triggered observation and control of linear systems via impulsive
  observers,'' {\em Journal of the Franklin Institute}, vol.~354, no.~1,
  pp.~372--391, 2017.

\bibitem{etienne2016event}
L.~Etienne and S.~Di~Gennaro, ``Event-triggered observation of nonlinear
  {L}ipschitz systems via impulsive observers,'' {\em IFAC-PapersOnLine},
  vol.~49, no.~18, pp.~666--671, 2016.

\bibitem{sijs2012event}
J.~Sijs and M.~Lazar, ``Event based state estimation with time synchronous
  updates,'' {\em IEEE Transactions on Automatic Control}, vol.~57, no.~10,
  pp.~2650--2655, 2012.

\bibitem{sijs2013event}
J.~Sijs, B.~Noack, and U.~D. Hanebeck, ``Event-based state estimation with
  negative information,'' in {\em Proceedings of the 16th international
  conference on information fusion \textnormal{Istanbul, Turkey}},
  pp.~2192--2199, 2013.

\bibitem{girard2014dynamic}
A.~Girard, ``Dynamic triggering mechanisms for event-triggered control,'' {\em
  IEEE Transactions on Automatic Control}, vol.~60, no.~7, pp.~1992--1997,
  2014.

\bibitem{tanwani2015using}
A.~Tanwani, A.~Teel, and C.~Prieur, ``On using norm estimators for
  event-triggered control with dynamic output feedback,'' in {\em 54th IEEE
  Conference on Decision and Control, \textnormal{Osaka, Japan}},
  pp.~5500--5505, 2015.

\bibitem{tabuada2007event}
P.~Tabuada, ``Event-triggered real-time scheduling of stabilizing control
  tasks,'' {\em IEEE Transactions on Automatic Control}, vol.~52, no.~9,
  pp.~1680--1685, 2007.

\bibitem{goedel2012hybrid1}
R.~Goebel, R.~G. Sanfelice, and A.~R. Teel, {\em Hybrid Dynamical Systems:
  Modeling, Stability, and Robustness}.
\newblock New Jersey, U.S.A: Princeton University Press, 2012.

\bibitem{cai2009characterizations}
C.~Cai and A.~R. Teel, ``Characterizations of input-to-state stability for
  hybrid systems,'' {\em Systems \& Control Letters}, vol.~58, no.~1,
  pp.~47--53, 2009.

\bibitem{abdelrahim2017robust}
M.~Abdelrahim, R.~Postoyan, J.~Daafouz, and D.~Ne{\v{s}}i{\'c}, ``Robust
  event-triggered output feedback controllers for nonlinear systems,'' {\em
  Automatica}, vol.~75, pp.~96--108, 2017.

\bibitem{dolk2016output}
V.~Dolk, D.~P. Borgers, and W.~P. M.~H. Heemels, ``Output-based and
  decentralized dynamic event-triggered control with guaranteed
  $\mathcal{L}_{p}$-gain performance and {Z}eno-freeness,'' {\em IEEE
  Transactions on Automatic Control}, vol.~62, no.~1, pp.~34--49, 2016.

\bibitem{liu2015event}
T.~Liu and Z.-P. Jiang, ``Event-based control of nonlinear systems with partial
  state and output feedback,'' {\em Automatica}, vol.~53, pp.~10--22, 2015.

\bibitem{luenberger1966observers}
D.~Luenberger, ``Observers for multivariable systems,'' {\em IEEE Transactions
  on Automatic Control}, vol.~11, no.~2, pp.~190--197, 1966.

\bibitem{krichman2001input}
M.~Krichman, E.~D. Sontag, and Y.~Wang, ``Input-output-to-state stability,''
  {\em SIAM Journal on Control and Optimization}, vol.~39, no.~6,
  pp.~1874--1928, 2001.

\bibitem{khalil2002nonlinear}
H.~K. Khalil, {\em Nonlinear systems}, vol.~3.
\newblock Prentice hall Upper Saddle River, NJ, 2002.

\bibitem{borgers2014event}
D.~P. Borgers and W.~P. M.~H. Heemels, ``Event-separation properties of
  event-triggered control systems,'' {\em IEEE Transactions on Automatic
  Control}, vol.~59, no.~10, pp.~2644--2656, 2014.

\bibitem{postoyan2014framework}
R.~Postoyan, P.~Tabuada, D.~Ne{\v{s}}i{\'c}, and A.~Anta, ``A framework for the
  event-triggered stabilization of nonlinear systems,'' {\em IEEE Transactions
  on Automatic Control}, vol.~60, no.~4, pp.~982--996, 2014.

\bibitem{he2012comparison}
H.~He, R.~Xiong, H.~Guo, and S.~Li, ``Comparison study on the battery models
  used for the energy management of batteries in electric vehicles,'' {\em
  Energy Conversion and Management}, vol.~64, pp.~113--121, 2012.

\bibitem{astolfi2021stubborn}
D.~Astolfi, A.~Alessandri, and L.~Zaccarian, ``Stubborn and dead-zone redesign
  for nonlinear observers and filters,'' {\em IEEE Transactions on Automatic
  Control}, vol.~66, no.~2, pp.~667--682, 2021.

\end{thebibliography}
\end{document}